\documentstyle[aps,floats,psfig]{revtex}

\begin{document}
\psfigurepath{.:plot:figure}
\twocolumn[\hsize\textwidth\columnwidth\hsize\csname @twocolumnfalse\endcsname
\bibliographystyle{unsrt}
\preprint{draft:thrd5.tex, \today}
\title{
Anisotropic three-dimensional magnetic fluctuations 
in heavy fermion CeRhIn$_5$
}
\author{Wei Bao$^1$, G. Aeppli$^2$, J. W. Lynn$^3$, 
P.G. Pagliuso$^1$, J.L. Sarrao$^1$, M. F. Hundley$^1$,
J.D. Thompson$^1$ and Z. Fisk$^{1,4}$}
\address{$^1$Los Alamos National Laboratory, Los Alamos, NM 87545}
\address{$^2$NEC Research, 1 Independence Way, Princeton, NJ 08540}
\address{$^3$Center for Neutron Research, National Institute of Standards
and Technology, Gaithersburg, MD 20899}
\address{$^4$Florida State University, Tallahassee, FL 32306}
\date{\today}
\maketitle
\begin{abstract}
CeRhIn$_5$ is a heavy fermion antiferromagnet that orders at 3.8~K. 
The observation of pressure-induced superconductivity
in CeRhIn$_5$ at a very high $T_C$ of 2.1~K for heavy fermion materials
has led to speculations regarding to its magnetic fluctuation spectrum.
Using
magnetic neutron scattering, we report anisotropic 
three-dimensional antiferromagnetic fluctuations with an
energy scale of less than 1.7~meV for temperatures as high as 3$T_C$. In
addition, the effect of the magnetic fluctuations on electrical
resistivity is
well described by the Born approximation.
\end{abstract}
\vskip2pc]

\narrowtext

The discovery of superconductivity in heavy fermion materials 
by Steglich et al.\cite{steg} just over two
decades ago generated much of our current thinking on the relation
between magnetism and superconductivity\cite{ott_fisk}. 
In the intervening years these insights have been extended
from the three-dimensional (3D) cerium and uranium-based
heavy fermion materials,
where the physics is thought to follow from the competition between the
single site Kondo effect and the RKKY interaction between spins on
different sites, to the layered cuprates in which the superconductivity
and magnetism are believed to derive from a two-dimensional (2D)
phenomenon.
Anisotropic superconductivity arising via the exchange of
antiferromagnetic
spin fluctuations has been investigated theoretically\cite{msrv}, 
and it is predicted that
2D antiferromagnetic spin fluctuations are superior to
3D fluctuations in elevating the superconducting transition
temperatures\cite{pmggl}.

Very recently, a family of heavy fermion compounds 
with chemical formula Ce$M$In$_5$ ($M=$ Rh, Ir, Co) has been
discovered\cite{hegger,joeIr,joeCo}. 
These materials, with the tetragonal HoCoGa$_5$ structure\cite{russ}
(space group \#123, P4/mmm),
consist of alternating layers of the cubic heavy fermion
antiferromagnet CeIn$_3$ and the transition metal complex $M$In$_2$. 
They display antiferromagnetism and
superconductivity in close proximity. Their superconducting transition
temperatures, $T_C$, are very high for heavy fermion systems. 
For example, $T_C$ is 2.1~K for CeRhIn$_5$ at 16 kbar\cite{hegger}, 
which is more than half of its N\'{e}el temperature at ambient pressure,
while the maximum $T_C=0.2$~K for cubic CeIn$_3$ at 25 kbar\cite{cmbr} 
is only 2\% of its $T_N$ at ambient pressure.
The $T_C$ of the ambient pressure
superconductor, CeCoIn$_5$, $T_C=2.3$~K, is the highest among all
heavy fermion superconductors\cite{joeCo}.
The enhancement of $T_C$ in layered Ce$M$In$_5$ over CeIn$_3$
has been suggested to be due to the quasi-2D structure of the new
materials, taking advantage of favorable coupling of 
2-D antiferromagnetic
fluctuations\cite{hegger,joeIr,joeCo,nqr,new1,haga}.
Here we describe an inelastic neutron scattering and electrical
transport study of the magnetic fluctuations for CeRhIn$_5$ in the
vicinity of $T_N$ which demonstrate that this hypothesis, in its
simplest form, does not seem applicable.

CeRhIn$_5$ ($a=4.652 \AA$, $c=7.542\AA$ at 295~K) 
is an incommensurate antiferromagnet below
$T_N=3.8$~K\cite{hegger,nqr,bao00a}. The small magnetic moments of the 
Ce ions, 0.26$\mu_B$ at 1.4~K, form a helical spiral along the
$c$ axis and are antiparallel for nearest-neighbor pairs in
the tetragonal basal plane, resulting in magnetic Bragg peaks
below $T_N$ at ${\bf q_M}=(m/2,n/2,l\pm \delta$) with $m$ 
and $n$ odd integers, $l$ integer and $\delta=0.297$\cite{bao00a}. 
If CeRhIn$_5$ is magnetically 3-D, magnetic fluctuations
are expected to be enhanced in the vicinity of ${\bf q_M}$ near $T_N$.
On the other hand, if CeRhIn$_5$ is 2-D, magnetic fluctuations
are enhanced in rods which go through magnetic Bragg points along
the $c^*$ axis of the same $m$ and $n$ indices.

Although single crystal samples of CeRhIn$_5$ with 
cm$^3$ size can be readily 
grown in our laboratory from an In flux\cite{hegger}, 
the high slow-neutron absorption coefficients of In and
Rh forced us to employ a thin plate-like sample. To reduce absorption
effects,
we also used neutrons of incident energy $E_i=35$~meV, selected with a
pyrolytic graphite (PG)(002) monochromator, 
in our neutron scattering experiments at NIST. The instantaneous
magnetic correlation function, ${\cal S}({\bf q})$,
is usually
measured with the two-axis method\cite{jens}.
However, for technical reasons, we chose the three-axis method for most
of this study. With PG (002) analyzers and horizontal collimations
40-40-40-40 and 60-40-40-open at the  
thermal neutron triple-axis spectrometers BT9 and BT2, respectively, the
energy window, given by   
the half-width-at-half-maximum (HWHM) of incoherent scattering, 
is 1.7~meV. 
As will be shown below, this is much
wider than the energy scale of magnetic fluctuations in CeRhIn$_5$
for the temperature range of interest. 
This three-axis configuration offers a better
signal-to-noise ratio than the two-axis method for the correlated
magnetic fluctuations in CeRhIn$_5$.
No sample-angle dependent absorption is observed in our data.
PG filters of 4 or 5 cm thickness were inserted in the incident
neutron beam to remove higher order
neutrons. The sample temperature was regulated using a heater installed
in our top-loading pumped
He cryostat. 

\begin{figure}[bt]
\centerline{
\psfig{file=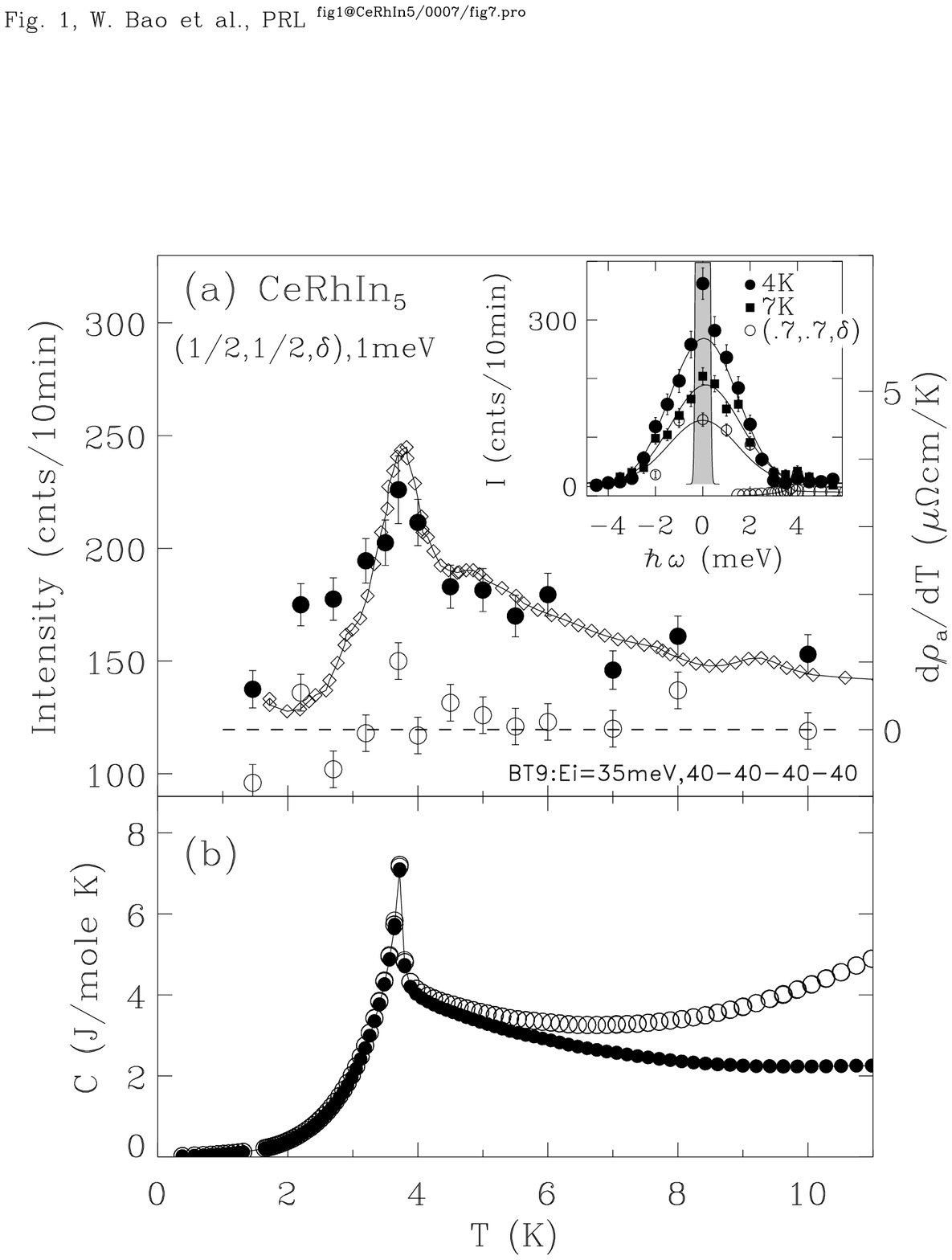,width=.75\columnwidth,angle=0,clip=}}
\caption{(a) Intensity of magnetic fluctuations,
${\cal S}({\bf q})-\langle M\rangle^2$,
at {\bf q}=(1/2,1/2,0.297) (solid circles)
as a function of temperature. 
Neutron scattering intensity measured at (0.7,0.7,0.297) (open circles)
indicates the background.
Temperature derivative of resistivity (open diamonds, scale on the
right), 
measured with current along the $a$ axis, 
closely follows magnetic fluctuations. 
Inset: Constant {\bf q}=(1/2,1/2,0.297) scans at 7K, 4 K (solid
symbols) and 1.7 K (shaded region). 
The constant {\bf q}=(0.7,0.7,0.297) scan
(open circles) measures incoherent neutron scattering. 
(b) Total specific heat (open circles) and magnetic specific heat
(solid circles) as a function of temperature.}
\label{fig1}
\end{figure}

The inset to Fig.~\ref{fig1}(a) shows energy scans at the magnetic Bragg
point (1/2,1/2,0.297).
Solid circles and squares
measure intensity of magnetic fluctuations at 4 and 7~K, respectively,
and the shaded
region represents the resolution-limited Bragg peak measured at 1.7~K.
Open circles indicate incoherent nuclear scattering measured away from
magnetic Bragg points. The width of the incoherent scattering, 
1.7 meV HWHM, is the 
energy window for magnetic fluctuations and is the same
as that of the scans at 4 and 7~K. This implies that the energy scale
for 
magnetic fluctuations, at least up to 7 K, is significantly
smaller than 1.7meV, and therefore a good
energy integration of the magnetic excitation spectrum was achieved with
our spectrometer configuration. 
In other words, the quasielastic condition of the two-axis method
was realized in our experiments\cite{jens}, and the
intensity of magnetic neutron scattering
measures the {\it instantaneous} magnetic correlation 
function ${\cal S}({\bf q})$\cite{neut_squire}:
\begin{equation}
I\sim \left|f({\bf q})\right|^2 
	\sum_{\mu,\nu}(\delta_{\mu\nu}
	-\widehat{\rm q}_{\mu}\widehat{\rm q}_{\nu})
	{\cal S}^{\mu\nu}({\bf q}),
\label{eq1}
\end{equation}
where $f({\bf q})$ is the magnetic form factor for a Ce$^{3+}$ ion, and
\begin{eqnarray}
	{\cal S}^{\mu\nu}({\bf q})&=&\hbar \int d\omega 
{\cal S}^{\mu\nu}({\bf q},\omega) \nonumber \\
&=&\sum_{{\bf R}\neq{\bf 0}} e^{i{\bf q}\cdot{\bf R}}
\langle {\rm M}_{\bf 0}^{\mu}(t) {\rm M}_{\bf R}^{\nu}(t) \rangle.
\end{eqnarray}
Here $\langle \dots \rangle$ denotes the thermodynamic average, and
${\rm M}_{\bf R}^{\mu}(t)$ is the $\mu$th Cartesian component
of the magnetic moment of the Ce ion
at position {\bf R} at time $t$.

Solid circles in Fig.~\ref{fig1}(a) are the temperature-dependent
intensity,
$I$, of neutrons scattered by magnetic fluctuations. They were measured
at the magnetic Bragg point (1/2,1/2,0.297) and 
with a finite energy transfer of 1 meV. 
This is larger than the elastic energy resolution so as to avoid the 
Bragg peak intensity below $T_N=3.8$~K, but well
inside the energy window for magnetic excitations [refer to the
inset in Fig.~\ref{fig1}(a)]. The peak at $T_N$
reflects the divergent magnetic fluctuations, 
${\cal S}({\bf q_M})-\langle M\rangle^2$, at a continuous phase
transition, 
rounded by the finite resolution of the spectrometer. 
Open circles were measured at 1 meV and (0.7,0.7,0.297) 
which is away from any magnetic Bragg points. The contrast between the
T-independent intensity here, which serves as a measure of the
background, and the strongly T-dependent signal at (1/2,1/2,0.297) 
reflects the strong spatial modulation of ${\cal S}({\bf q})$. 

Magnetic resistivity, or angular integration of intensity of 
conduction electrons scattered by magnetic fluctuations, 
has been calculated by Fisher and Langer\cite{fisherl}
using the Born approximation. The long mean-free-path
for conduction electrons observed in our sample, 
$l/a \sim 10^2$ near $T_N$, estimated from the
measured resistivity [see Fig.~\ref{fig3}(b)] and Hall
coefficient\cite{mike},
implies that the anomalies in $d\rho/dT$ and in magnetic specific heat
at $T_N$ are directly related to the singularity in coherent
scattering of conduction electrons by 
magnetic fluctuations, ${\cal S}({\bf q})-\langle M\rangle^2$.
Line-connected diamonds in Fig.~\ref{fig1}(a) are experimental
$d\rho_a/dT$, and
solid circles in Fig.~\ref{fig1}(b) are magnetic specific heat, 
obtained by subtracting specific heat of LaRhIn$_5$ from the total
specific heat of CeRhIn$_5$ (open circles)\cite{hegger}.
The similarity among these quantities indicates the 
prominence of the Fisher-Langer
mechanism in CeRhIn$_5$ at low temperature. This behavior has previously
been observed in ferromagnetic metals, such as Ni\cite{fl_Ni}, and
antiferromagnets, such as PrB$_6$\cite{fl_AF}.
However, it is astonishing that
the Born approximation is sufficient to account for the influence of
magnetic fluctuations
on the electrical transport in this heavy fermion material. 

\begin{figure}[bt]
\centerline{
\psfig{file=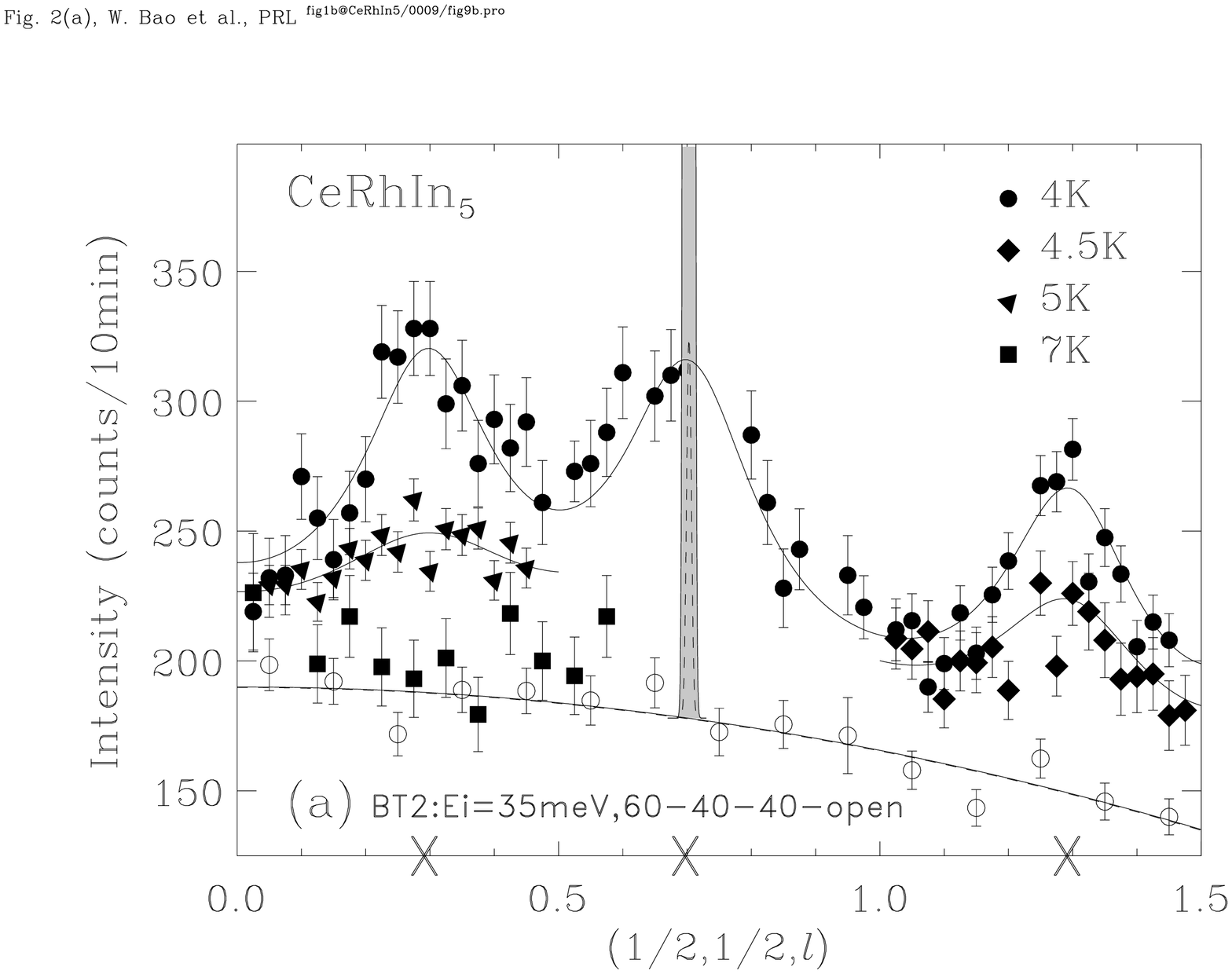,width=.7\columnwidth,angle=90,clip=}}
\vskip -4mm
\centerline{
\psfig{file=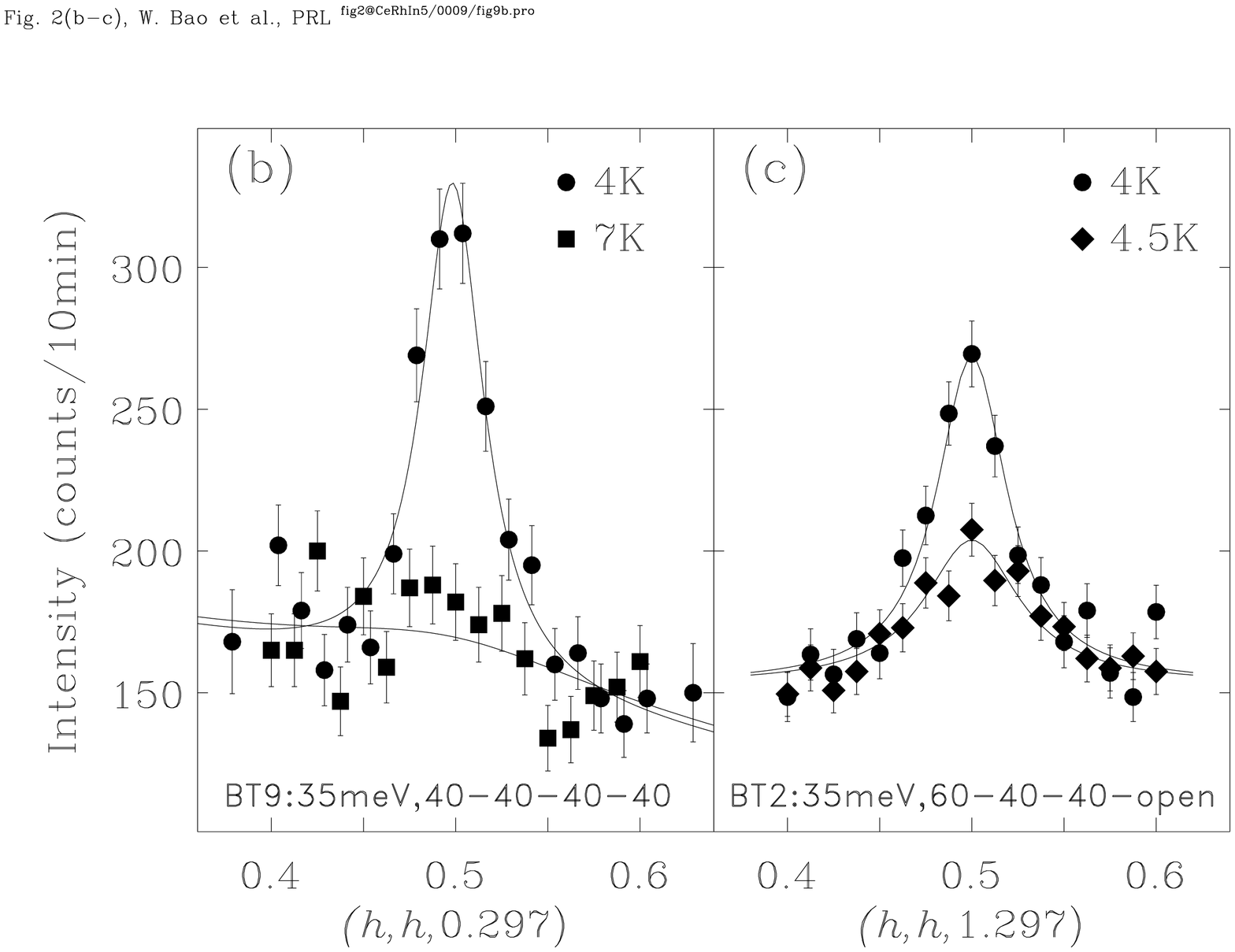,width=.7\columnwidth,angle=90,clip=}}
\caption{Instantaneous magnetic correlation function,
${\cal S}({\bf q})$, for (a) ${\rm q}=(1/2,1/2,l$), and for {\bf q}
along the (110) direction at (b) $l=0.297$ and (c) $l=1.297$
at various temperatures. 
Open circles in (a) are background measured along (0.4,0.4,$l$). 
Crosses indicate the magnetic
Bragg peak positions below $T_N=3.8$~K.
The shaded peak is the magnetic Bragg peak measured at 1.7~K
and the dashed line represents its intensity divided by 120.}
\label{fig2}
\end{figure}

\begin{figure}[bt]
\centerline{
\psfig{file=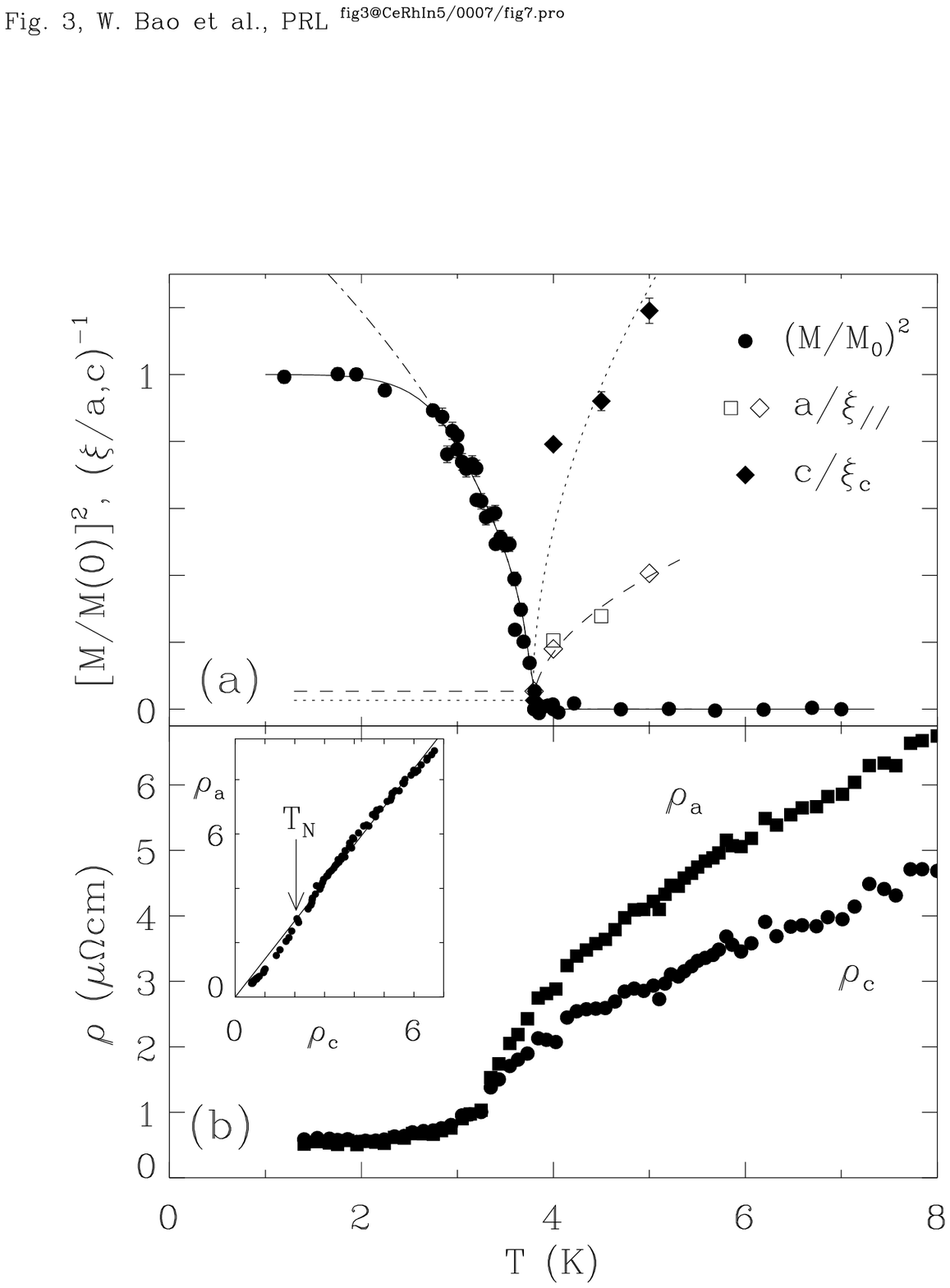,width=.7\columnwidth,angle=0,clip=}}
\vskip -6mm
\caption{(a) Solid circles represent the square of the order parameter.
The dot-dashed line is $(1-T/T_N)^{0.5}$.
Open symbols and solid diamonds are inverse magnetic correlation 
lengths in the tetragonal
basal plane and along the $c$ axis respectively.
The dashed and dotted line are guides to the eye above $T_N$ and
represent the Bragg peak width below $T_N$.
(b) Resistivity, measured with current parallel to the $a$ axis
(squares) and the $c$ axis (circles) respectively, as a function of 
temperature. 
Inset: $\rho_c$ vs $\rho_a$ with temperature as the implicit
variable. The arrow indicates the N\'{e}el temperature.  }
\label{fig3}
\end{figure}

We now turn to the spatial dependence of magnetic correlations 
in CeRhIn$_5$. A survey of 
${\cal S}({\bf q})$ along the $c$ axis and in tetragonal
basal plane intersecting Bragg points (1/2,1/2,$\delta$) and
(1/2,1/2,$1+\delta$) is shown in Fig.~\ref{fig2}.
The solid circles were measured at 4~K. Strong modulation of 
${\cal S}({\bf q})$ with peaks at magnetic Bragg points,
in scans along both the $c$ axis [Fig.~\ref{fig2}(a)] and
in the basal plane [Fig.~\ref{fig2}(b) and (c)], is apparent, 
indicating 3-D magnetic fluctuations. 
Fitting the data to an infinite sum of
Lorentzians, centered at (1/2,1/2,$l\pm \delta$), we obtain magnetic
correlation lengths at 4~K of $\xi_c=9.5(1) \AA$ along the $c$ axis 
and $\xi_{\parallel}=23(1) \AA$ in the basal plane.
While $\xi_{\parallel}$ is about 5 times the nearest-neighbor
distance, $a$, of Ce ions in the basal plane,
$\xi_c$ is only 1.3 times the interplane Ce distance, $c$.
However, this is different from a 2-D magnetic system, 
in which the intraplanar
magnetic correlation length is orders of magnitude longer than the
interplanar magnetic correlation length.

With increasing temperature (diamonds at 4.5~K, triangles at 5~K
and squares at 7~K in Fig.~\ref{fig2}), 
the magnetic peak intensity and correlation lengths are quickly reduced.
At 7~K, little modulation in ${\cal S}({\bf q})$ can be detected
either along the $c$ axis [Fig.~\ref{fig2}(a)] or in the basal plane
[Fig.~\ref{fig2}(b)], indicating the loss of magnetic correlations
above 7~K.
Temperature dependences of inverse $\xi_{\parallel}$ and $\xi_c$
in units of their respective inter-Ce distances are shown
in Fig.~\ref{fig3}(a). That they vary with T in the same way 
implies that the magnetic
system in CeRhIn$_5$, although anisotropic, behaves three-dimensionally
below 7~K, and no 3-D to 2-D crossover is found prior to the complete
disappearance of intersite magnetic correlations.

\begin{figure}[bt]
\centerline{
\psfig{file=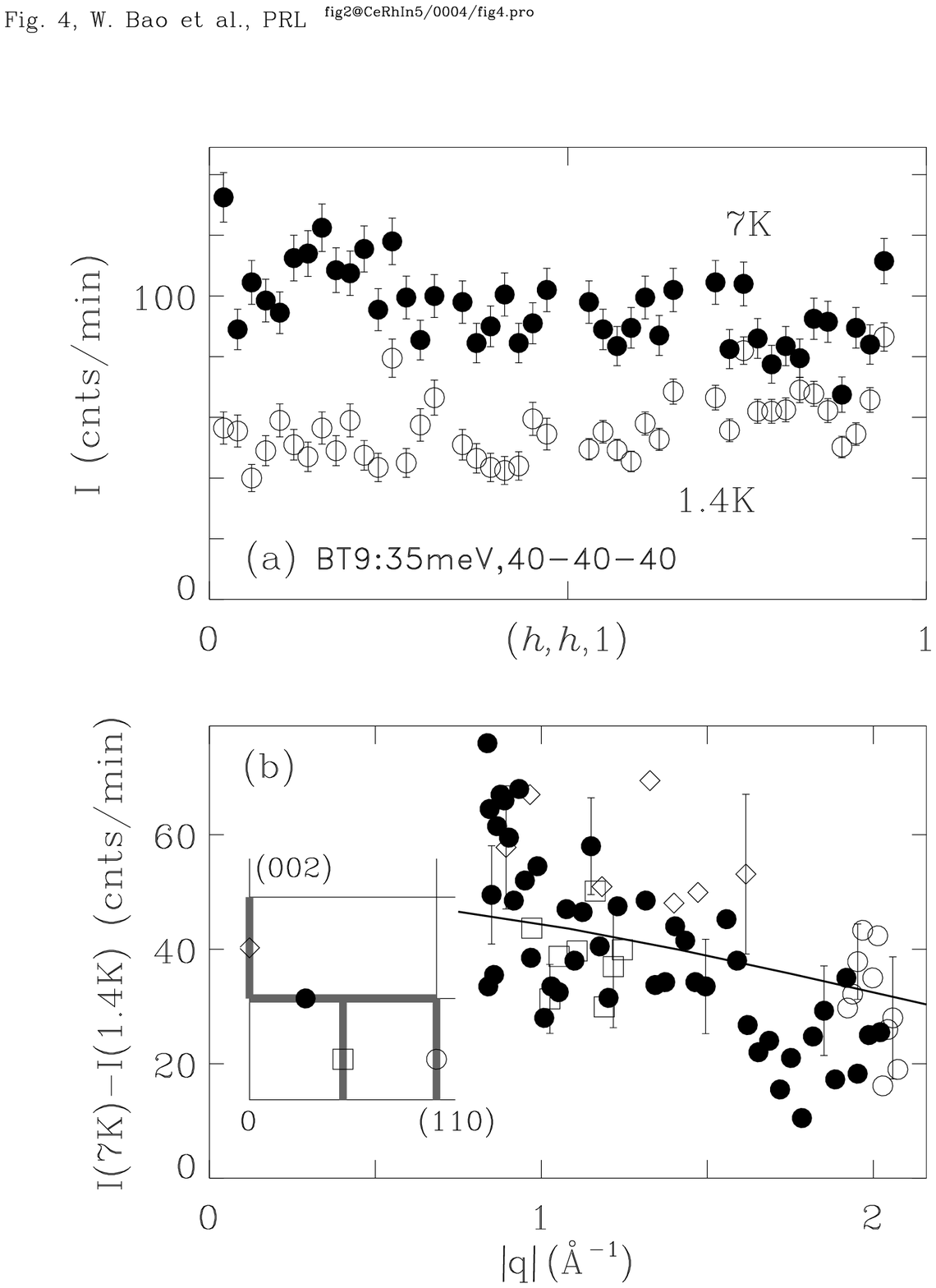,width=.7\columnwidth,angle=0,clip=}}
\caption{(a) Quasielastic scan along ($h$,$h$,1) at
1.4 and 7~K. The extra intensity at 7 K is due to magnetic fluctuations.
(b) Magnetic intensity at 7 K as a function of $|{\bf q}|$. 
Solid circles were from the ($h,h,$1) scan covering $h=0$-1, 
open diamonds from the (0,0,$l$) scan with $l=1$-2, 
open squares from the (0.5,0.5,$l$) scan and 
open circles from the (1,1,$l$) scan with $l=0$-1, as shown 
by the shaded lines in the inset.}
\label{fig4}
\end{figure}

To further examine the magnetic state at 7~K, we used the two-axis
method
to measure magnetic scattering in a
wider range in reciprocal space than the scans shown
in Fig.\ref{fig2} [refer to the inset to Fig.~\ref{fig4}(b)]. 
Fig.~\ref{fig4}(a) presents a quasielastic scan along ${\bf q}=(h,h,1)$
for $h$ ranging from 0 to 1. The extra intensity at 7~K derives
from fluctuations of magnetic moments which condense into Bragg peaks
at 1.4~K. Any multi-phonon contribution to the extra
flat intensity should be minimal in view of the low temperatures and
small wave numbers. Magnetic intensity at 7~K obtained in various
scans with different symmetries is plotted in Fig.~\ref{fig4}(b)
as a function of $q$. The solid line through the data points is the
square of the Ce$^{3+}$ form factor\cite{formf_ce}.
We notice that (i) polarization dependence is negligible, 
and (ii) there is little $q$ dependence in magnetic intensity
beyond the form factor. This means that at 7K, the 
magnetic moments in CeRhIn$_5$ fluctuate independently.
The local anisotropic magnetic field, which aligns magnetic moments
in the basal plane in the ordered state, has lost its effect on
the moments at 7K. This temperature is also the locus of the peak in the
bulk magnetic susceptibility\cite{hegger}, which,
thus, is related to development of short-range antiferromagnetic
correlations.
A similar correspondence between antiferromagnetic short-range order and
the susceptibility maximum exists also in the heavy fermion
superconductor
UPt$_3$\cite{agold}, which orders antiferromagnetically below 5~K with a
tiny magnetic moment 0.02(1)$\mu_B$ per U\cite{gabe_upt3}.

Even though from the point of view of statistical physics, CeRhIn$_5$ is
clearly 3-D, at the microscopic level it still seems
quite 2-D: the lattice
structure is tetragonal with $c/a>1$, and the anisotropic correlation
length together with the pairwise appearance of magnetic Bragg peaks
concentrate magnetic spectral weight in reciprocal space
in ellipsoids which are elongated in the $c$ direction.  While this
still differs from the more ideally 2-D situation for the cuprate
superconductors, the resemblance may be sufficient that phase-space
arguments\cite{pmggl,joeCo}, favoring 2-D over 3-D superconductivity
when the mediating bosons are antiferromagnetic fluctuations, may still
be responsible for the unusually high superconducting $T_C$'s of the
layered compounds based on CeIn$_3$. One classic way in which to decide
the nature of the coupling to the bosons responsible for
superconductivity is to examine the electrical resistivity $\rho$ above
$T_C$. We have already shown that the T-dependence of $\rho$ at low T is due to
scattering from the same magnetic fluctuations which would be needed to
produce superconductivity. Should the substantial anisotropy of the
fluctuations matter for the superconductivity, it would also appear in
the resistance anisotropy above $T_C$. We have therefore measured the
electrical resistivity for our samples using the standard 4-probe method,
with currents along the $a$
and $c$ axes [see Fig.~\ref{fig3}(b)]. The first point to
notice is that in contrast to naive expectations and what is
seen in the cuprates, $\rho_c$ is somewhat smaller than $\rho_a$
in this temperature range. This
does mean, however, that the electrons scatter more, but not much more,
strongly from in-plane rather than out-of-plane spin fluctuations.
Second, $\rho_c$ and $\rho_a$ have very similar temperature dependences.
The proportionality between them may be better seen in the inset to
Fig.~\ref{fig3}(b)
and is very strict down to  a temperature slightly above $T_N$. 
This expands our conclusion from the neutron scattering
measurements to assert that as far as the conduction electrons and
simple antiferromagnetic mechanisms for superconductivity are concerned,
the magnetic correlations in CeRhIn$_5$ are 3-D.

In summary, we observe the development of {\bf q} dependent magnetic 
correlations below 7~K in CeRhIn$_5$. Although anisotropic, 
these correlations are 3-D in nature and 
have a characteristic energy less than 1.7~meV.
Conduction electron scattering in the vicinity of $T_N$ is dominated by
these
antiferromagnetic fluctuations, and in a manner consistent with weak
coupling (the Born approximation is valid) between renormalized, heavy
conduction electrons and the $f$ magnetic moments  
of the Ce ions uncompensated by the Kondo process. The electrical
resistivity and magnetic fluctuations are sufficiently isotropic that it
is unlikely that a specifically 2-D mechanism involving
antiferromagnetic fluctuations is the explanation for the unusual
superconductivity of the CeIn$_3$-based layered compounds, 
unless these effects are strongly compound or pressure dependent.

We thank J.M. Lawrence for valuable discussion.
Work at Los Alamos was performed under the auspices of the US Department
of Energy. ZF gratefully acknowledges NSF support at FSU. PGP
acknowledges also
FAPESP-BR.

\end{document}